\documentclass[twocolumn,showpacs,prl]{revtex4}
\usepackage{bm}
\usepackage{epsfig}
\newcommand{\nix}[1]{}

\begin{document}

\title{
Experimental Separation of Rashba and Dresselhaus Spin-Splittings
\\in Semiconductor Quantum Wells }
\author{S.D.~Ganichev$^{1,2}$, V.V.~Bel'kov$^2$, L.E.~Golub$^2$, E.L.~Ivchenko$^2$,
Petra~Schneider$^1$, S.~Giglberger$^1$, J.~Eroms$^1$, J.~DeBoeck$^3$, G.~Borghs$^3$, W.~Wegscheider$^1$,
D.~Weiss$^1$, W.~Prettl$^1$}
\affiliation{$^1$Fakult\"{a}t Physik, University of Regensburg,
93040, Regensburg, Germany}
\affiliation{$^2$A.F.~Ioffe Physico-Technical Institute, Russian
Academy of Sciences, 194021 St.~Petersburg, Russia}
\affiliation{$^3$IMEC, Kapeldreef 75, B-3001 Leuven, Belgium}


\begin{abstract}
The relative strengths of Rashba and Dresselhaus terms describing
the spin-orbit coupling in semiconductor quantum well (QW)
structures are extracted from photocurrent measurements on $n$-
type InAs QWs containing a two-dimensional electron gas (2DEG).
This novel technique makes use of the angular distribution of the
spin-galvanic effect at certain directions of spin orientation in
the plane of a QW. The ratio of the relevant Rashba and
Dresselhaus coefficients can be deduced directly from experiment
and does not relay on theoretically obtained quantities. Thus our
experiments open a new way to determine the different
contributions to spin-orbit coupling.
\end{abstract}

\maketitle

\begin{figure}
\centerline{\epsfysize 70mm \epsfbox{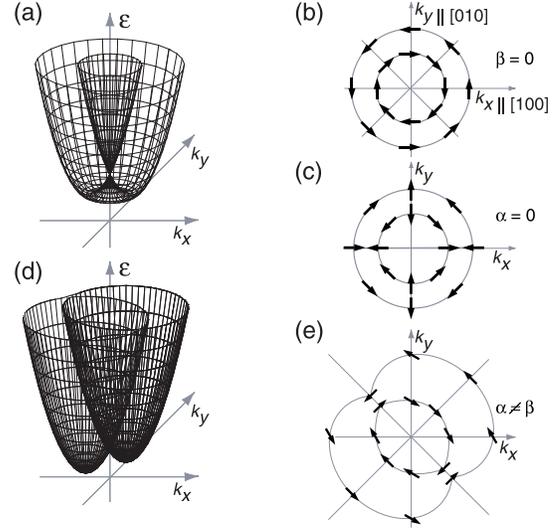}}
\caption{Schematic 2D  band structure with $\bm{k}$-linear terms
for C$_{2v}$ symmetry for different relative strengths of SIA and
BIA and the distribution of spin orientations at the 2D Fermi
energy. (a) shows the case of only Rashba or Dresselhaus
spin-orbit coupling and (d) represents the case of simultaneous
presence of both contributions. Arrows indicate the orientation of
spins. } \label{fig1}
\end{figure}

The manipulation of the spin of charge carriers  in semiconductors
is one of the key problems in the field of spintronics (see e.g.
\cite{spintronicbook02}). In the paradigmatic spin transistor,
e.g. proposed by Datta and Das~\cite{Datta1990p665}, the electron
spins, injected from a ferromagnetic contact into a
two-dimensional electron system are controllably rotated during
their passage from source to drain by means of the Rashba
spin-orbit coupling~\cite{Bychkov84p78}. The  coefficient
$\alpha$, which describes the strength of the Rashba spin-orbit
coupling, and hence the degree of rotation, can be tuned by gate
voltages. This coupling stems from the inversion asymmetry of the
confining potential of two-dimensional electron (or hole) systems.
The dependence of  $\alpha$ on the gate voltage has been shown
experimentally by analyzing the beating pattern observed in
Shubnikov-de Haas (SdH)
oscillations~\cite{Das89p1411,Luo90p7685,Nitta97p1335,Engels97pr1958,Heida98p11911,Hu99p7736,Grundler00p6074}.
In addition to the Rashba coupling, caused by structure inversion
asymmetry (SIA),  also a Dresselhaus type of coupling contributes
to the spin-orbit interaction. The latter is due to bulk inversion
asymmetry (BIA)~\cite{Roessler1989p376,Dyakonov86p110} and the
interface inversion asymmetry
(IIA)~\cite{Vervoort97p12744,Roessler02p313}. The BIA and IIA
contributions are phenomenologically inseparable and described
below by the generalized Dresselhaus parameter $\beta$. Both,
Rashba and Dresselhaus couplings, result in spin-splitting of the
band (Fig.~\ref{fig1}) and give rise to a variety of spin
dependent phenomena which allow to evaluate the magnitude of the
total spin splitting of electron subbands.

However, usually it is  not possible to extract the  re\-lative
contributions of Rashba and Dresselhaus terms to the spin-orbit
coupling. To obtain the Rashba coefficient $\alpha$,
 the Dresselhaus contribution is normally
neglected~\cite{Luo90p7685,Nitta97p1335,Engels97pr1958,Heida98p11911,Hu99p7736,Grundler00p6074}.
At the same time, Dresselhaus and Rashba terms can interfere in
such a way that macroscopic effects vanish though the indivi\-dual
terms are large~\cite{Averkiev02pR271,Tarasenko02p552}. For
example, both terms can cancel each other resulting in a vanishing
spin splitting in certain {\boldmath$k$}-space
directions~\cite{review2003spin}. This cancellation leads to the
disappearance of an anti-localization~\cite{Knap1996p3912}, the
absence of spin relaxation in specific crystallographic
directions~\cite{Averkiev1999p15582,Averkiev02pR271}, and the lack
of SdH beating ~\cite{Tarasenko02p552}. In a recent
paper~\cite{Schliemann03p146801} the importance of both Rashba and
Dresselhaus terms was pointed out: tuning $\alpha$ such that
$\alpha = \beta$ holds, allows to build a non-ballistic spin-field
effect transistor.

Below we demonstrate that angular dependent measurements of the
spin-galvanic photocurrent~\cite{Nature02} allow to separate
contributions due to Dresselhaus and Rashba terms. Here, we make
use of the fact that these terms contribute differently for
particular crystallographic directions. Hence, by mapping the
magnitude of the photocurrent in the plane of the QW the ratio of
both terms can be directly determined from experiment.

Before turning to experiment we briefly summarize the consequences
of Rashba and Dresselhaus terms on the electron dispersion and on
the spin orientation of the 2DEG's electronic states. We consider
QWs of zinc-blende structure grown in [001] direction. For the
corresponding C$_{2v}$ symmetry the spin-orbit part $\hat{H}_{SO}$
of the Hamiltonian $\hat{H} = \hbar^2 \bm{k}^2 / 2m^* +
\hat{H}_{SO}$ contains the 
Rashba term as well as the
Dresselhaus terms according to
\begin{equation}
\label{eq1}
\hat{H}_{SO} = 
 \alpha (\sigma _x k_y  - \sigma _y k_x
 ) + \beta( \sigma _x k_x  - \sigma _y  k_y )
\end{equation}
where {\boldmath$k$} is the electron wavevector, and
\mbox{\boldmath$\sigma$} is the vector of the Pauli matrices.
Here, the $x$-axis is aligned along the [100]-direction, $y$ along
[010], and $z$ is the growth direction (see Fig.\ref{fig1}). The
Hamiltonian of Eq.~(\ref{eq1}) contains only terms linear in ${\bm
k}$. As we show below terms cubic in ${\bm k}$ in our experiments
only change the strength of $\beta$ leaving the Hamiltonian
unchanged.

To illustrate the resulting energy dispersion in Fig.~\ref{fig1}
we plot the eigenvalues of $\hat{H}$,
$\varepsilon$({\boldmath$k$}), and contours of constant energy in
the $k_{x}$,$k_{y}$ plane for different ratios of $\alpha$ and
$\beta $. For $\alpha \ne 0,\beta = 0$ and $\alpha = 0,\beta \ne
0$ the dispersion has the same shape and consists of two shifted
parabolas in all directions, displayed in Fig.~\ref{fig1}a.
However, Rashba and Dresselhaus terms result in a different
pattern of the eigenstate's spin orientation in
{\boldmath$k$}-space. The distribution of this spin orientation
can be visualized by writing the spin-orbit interaction term in
the form $\hat{H}_{SO} = \mbox{\boldmath$\sigma$} \cdot
\mbox{\boldmath$B$}_{eff}(\bm{k})$ where
$\mbox{\boldmath$B$}_{eff}(\bm{k})$ is an effective magnetic field
which provides the relevant quantization axes~\cite{Silva92p1921}.
Obviously, as long as time-reversal is not broken, the
Kramers-relation $\varepsilon (\mbox{\boldmath$k$}, \uparrow ) =
\varepsilon(-\mbox{\boldmath$k$}, \downarrow $ )
 holds.
 By
comparison with Eq.~(\ref{eq1}) one obtains for pure Rashba
($\beta $=0) and pure Dresselhaus ($\alpha $=0) coupling the
corresponding effective magnetic fields,
$\mbox{\boldmath$B$}_{eff}^{(R)} = \alpha (k_y , - k_x )$ and
$\mbox{\boldmath$B$}_{eff}^{(D)} = \beta (k_x ,-k_y )$,
respectively. The spin orientations for Rashba and Dresselhaus
coupling are schema\-tically shown in Fig.~\ref{fig1}b
and~\ref{fig1}c by arrows. Here it is assumed that $\alpha > \beta
>0$. For the Rashba  case the effective magnetic field and
hence the spin is always perpendicular to the corresponding
{\boldmath$k$}-vector (Fig.~\ref{fig1}b). In contrast, for the
Dresselhaus contribution, the angle between {\boldmath$k$}-vector
and spin depends on the direction of {\boldmath$k$}. In the
presence of both  Rashba and  Dresselhaus spin-orbit couplings,
relevant for C$_{2v}$ symmetry, the $[1\bar {1}0]$ and the [110]
axes become strongly non-equivalent. For
$\mbox{\boldmath$k$}\parallel [1\bar {1}0]$ the eigenvalues of the
Hamiltonian are then given by $\varepsilon =
\hbar^2\mbox{\boldmath$k$}^2 / 2m^*\pm (\alpha - \beta
)\mid\mbox{\boldmath$k$}\mid$ and for
$\mbox{\boldmath$k$}\parallel [110]$ by $\varepsilon =
\hbar^2\mbox{\boldmath$k$}^2 / 2m^*\pm (\alpha + \beta
)\mid\mbox{\boldmath$k$}\mid$. This anisotropic dispersion
$\varepsilon$({\boldmath$k$})  is sketched in Fig.~\ref{fig1}d,
and the corresponding contours of constant energy together with
the spin orientation of selected {\boldmath$k$}-vectors are shown
in Fig.~\ref{fig1}e.

\begin{figure}
  \centerline{\epsfxsize 65mm \epsfbox{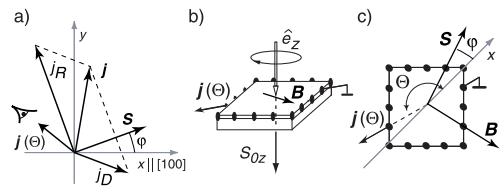}}
\caption{ Angular dependence of the spin-galvanic current (a) and
the geometry of the experiment (b) and (c). } \label{fig2}
\end{figure}

Angular dependent investigations of spin photocurrents  provide a
direct measure of the anisotropic orientation of spins in
{\boldmath$k$}-space and hence of the different contributions of
the Rashba and the Dresselhaus terms. We employ the spin-galvanic
effect to extract the ratio of the Rashba and the Dresselhaus
contributions. The spin-galvanic current is driven by the electron
in-plane average spin {\boldmath$S_\parallel$} according
to~\cite{Nature02,review2003spin}:
\begin{equation}
\label{eq2} \mbox{\boldmath$j$}_{SGE}  \propto \left(
{{\begin{array}{*{20}c}
 \beta \hfill & -\alpha \hfill \\
  \alpha  \hfill & - \beta \hfill \\
\end{array} }} \right)\mbox{\boldmath$S_\parallel$}
\end{equation}
%

Therefore, the spin galvanic current {\boldmath$j$}$_{SGE}$  for a
certain direction of \mbox{\boldmath$S_\parallel$} consists of
Rashba and Dresselhaus coupling induced currents,
$\mbox{\boldmath$j$}_R$ and $\mbox{\boldmath$j$}_D$ (see
Fig.~\ref{fig2}a). Their magnitudes are $j_R \propto \alpha
\left|\mbox{\boldmath$S_\parallel$}\right|$, $j_D \propto \beta
\left|\mbox{\boldmath$S_\parallel$}\right|$ and their ratio is
\begin{equation} \label{jj}
j_R/j_D = \alpha/\beta\:.
\end{equation}

For {\boldmath$S_\parallel$} oriented along one of the cubic axes
it follows from Eq.~(\ref{eq2}) that the currents flowing along
 and perpendi\-cular to {\boldmath$S_\parallel$} are equal
to $j_D$ and $j_R$, respectively, yielding experimental access to
determine $\alpha/\beta$.

 The experiments are  carried  out on
(001)-oriented $n$-type heterostructures having $C_{2v}$ point
symmetry. InAs/Al$_{0.3}$Ga$_{0.7}$Sb single QW of 15~nm width
with free carrier density of $1.29\cdot10^{12}$~cm$^{-2}$ and
mobility $2.05\cdot10^4$~cm$^2$/(Vs) at $T$=293~K  were grown by
molecular-beam-epitaxy. The sample edges are oriented along the
[1$\bar{1}$0] and [110] crystallographic axes. Eight pairs of
contacts allow to probe the photocurrent in different directions
(see Fig.~\ref{fig2}b). For  optical spin orientation we use a
high power pulsed molecular far-infrared  NH$_3$
laser~\cite{PhysicaB99tun}.  The linearly polarized radiation at a
wavelength 148\,$\mu$m with a power of 10\,kW is mo\-dified to
circularly polarized by using a $\lambda/4$ quartz  plate. The
photocurrent {\boldmath$j$}$_{SGE}$ is measured at room
temperature in unbiased structures via the voltage drop across a
50~$\Omega$ load resistor in a closed circuit
configuration~\cite{review2003spin}. It is detected for right
($\sigma_+$) and left ($\sigma_-$) handed circularly polarized
radiation. The spin-galvanic current $j_{SGE}$, studied here, is
extracted after eliminating current contributions which are
independent of the light polarization~\cite{Modena}:
$j_{SGE}=\left(j_{\sigma_+}-j_{\sigma_-}\right)/2$.

The non-equilibrium in-plane spin polarization
{\boldmath$S_\parallel$} is prepared as described
recently~\cite{Nature02}: Circularly polarized light at normal
incidence on the 2DEG plane, induces indirect (Drude-like)
electron transitions in the lowest conduction subband of our
$n$-type samples resul\-ting in a monopolar spin
orientation~\cite{PASPS02} in the $z$-direction
(Fig.~\ref{fig2}b).  An in-plane magnetic field ($B=1$\,T) rotates
the spin around the magnetic field axis (precession) and results
in a non-equilibrium in-plane spin polarization $S_\parallel
\propto \omega_L \tau_s$, where $\omega_L$ is the Larmor frequency
and $\tau_s$ is the spin relaxation time. The angle between the
ma\-gnetic field and {\boldmath$S_\parallel$} in general depends
on details of the spin relaxation process. In the InAs QW
structure investigated here, the isotropic Elliot-Yafet spin
relaxation mechanism
dominates~\cite{Takeuchi1999p318,Averkiev02pR271}. Thus the
in-plane spin polarization {\boldmath$S_\parallel$} of
photoexcited carriers is always perpendicular to {\boldmath$B$}
and can be varied by rotating {\boldmath$B$} around $z$ as
illustrated in Fig.~\ref{fig2}c. This excess spin polarization
{\boldmath$S_\parallel$} leads to an increase of the population of
the corresponding spin-polarized states. Due to asymmetric spin
relaxation an electric current results~\cite{Nature02}.

\begin{figure}
\centerline{\epsfxsize 64mm \epsfbox{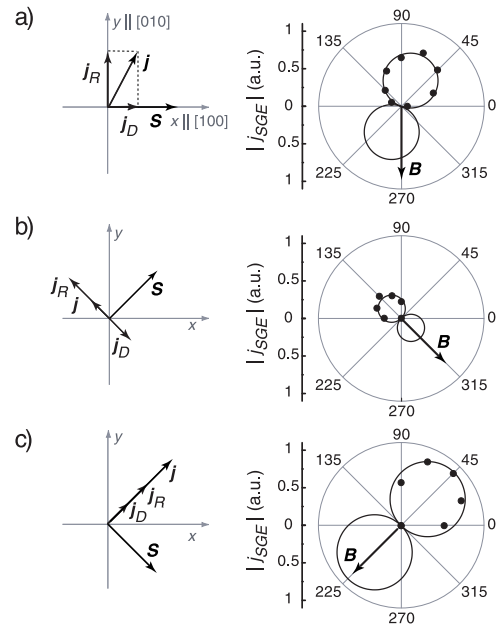}}
\caption{Photocurrent in $n$-type InAs single QWs. Left  plates
indicate three selected relations between spin polarization and
current contributions (after Eq.~\protect\ref{eq2}). Right plates
show measurements of the spin galvanic current as a function of
angle $\Theta$. Data are presented in polar coordinates.}
\label{fig3}
\end{figure}

To obtain the Rashba- and Dresselhaus contributions the
spin-galvanic effect is measured for a fixed orientation of
{\boldmath$S_\parallel$} for all accessible directions $\Theta$
(see Fig.~\ref{fig2}c). According to Eq.~(\ref{eq2}) the current
$\mbox{\boldmath$j$}_R$ always flows perpendicularly to the spin
polarization {\boldmath$S_\parallel$}, and {\boldmath$j$}$_D$
encloses an angle $- 2\varphi$ with {\boldmath$S_\parallel$}. Here
$\varphi$ is the angle between {\boldmath$S_\parallel$} and the
$x$-axis. Then, the current component along any direction given by
angle $\Theta$
 can be written as a sum of the projections of
$\mbox{\boldmath$j$}_R$ and $\mbox{\boldmath$j$}_D$ on this
direction
\begin{equation}
\label{eq3} j_{SGE} (\Theta ) = j_D\cos (\Theta + \varphi ) +
j_R\sin (\Theta - \varphi ).
\end{equation}
Three directions of spin po\-pulation {\boldmath$S_\parallel$} are
particularly suited to extract the ratio between Rashba  and
Dresselhaus terms. These geometries are sketched in
Fig.~\ref{fig3} (a-c), left column. In Fig.~\ref{fig3}a, the spin
polarization $\bm{S_\parallel}$ is set along [100] ($\varphi =
0$). Then from Eq.~(\ref{eq3}) follows that the currents along the
[100]-direction $(\Theta=0)$ and [010]-direction ($\Theta =
\pi/2$) are equal to $j_D$ and $j_R$, respectively, as shown on
the left hand side of Fig.~\ref{fig3}a. Fig.~\ref{fig3}b
illustrates another geometry. For a non-equilibrium spin
polarization induced along
 [110] ($\varphi=  \pi/4$) Eq.~(\ref{eq3}) predicts that  the
current has its maximum value $j  = j_R - j_D$ at $\Theta=3\pi/4$.
If the spin is aligned along [1$\bar{1}$0] ($\varphi= -\pi/4$ in
Fig.~\ref{fig3}c), on the other hand, the maximum current $j
=j_R+j_D$ is expected to flow under an angle of $\Theta=\pi/4$.
Thus, the relative strength of the measured $j_R-j_D$ and
$j_R+j_D$ values allows a straight forward determination of
$j_R/j_D=\alpha/\beta$.

The results are shown in Fig.~\ref{fig3}. The left hand side of
Fig.~\ref{fig3} displays the geometric arrangement discussed above
and shows the direction of the photogenerated spins
{\boldmath$S_\parallel$} and the resulting Rashba and Dresselhaus
currents. The corresponding experimentally obtained currents
measured in different directions are presented in polar
coordinates on the right hand side of the figure. The current's
magnitude is normalized to the maximum value of the spin-galvanic
current obtained if Rashba and Dresselhaus contributions point in
the same direction (Fig.~\ref{fig3}c). The ratio of Rashba and
Dresselhaus currents can be directly read off from the right hand
side of Fig.~\ref{fig3}a, $j_R/j_D=j(\pi/2)/j(0)$
 or can be evaluated from the maximum currents   $j$
 in Fig.~\ref{fig3}b and~\ref{fig3}c. Both procedures give the same result:
$j_R/j_D=2.15$. Moreover, all data on the right hand side of
Fig.~\ref{fig3} are in excellent agreement with the picture given
above: Using $\alpha/\beta=2.15$, the three sets of the data
points can be  fitted simultaneously by Eq.~(\ref{eq3}) without
additional fitting parameters.

The value of 2.15 agrees with theoretical
results~\cite{Lommer88p728} which predict a dominating Rashba
spin-orbit coupling for InAs QWs and is also consistent with
recent expe\-riments~\cite{Knap1996p3912,Nitta97p1335}. For InGaAs
QWs, having similar sample parameters as the devices investigated
here, $\alpha/\beta$ ratios were obtained from weak
antilocalization experiments~\cite{Knap1996p3912} and {\bf k
$\cdot$ p} calculations~\cite{Pfeffer99pr5312}. The corresponding
values ranged between 1.5~-~1.7 and 1.85, respectively. These
results are in a good agreement with our findings. The ratio of
Rashba and Dresselhaus terms has previously been estimated by
means of Raman spectroscopy~\cite{Jusserand95p4707} and transport
investigations~\cite{Knap1996p3912,Miller03p076807}. In contrast
to these works our method allows to measure directly the relative
strength of Rashba and Dresselhaus terms and does not require any
additional theoretical estimations.

So far we have not addressed the role of a contribution cubic in
${\bm k}$ in the Hamiltonian $\hat{H}_{SO}$. This results in terms
proportional to $k^3$ in the Hamiltonian which vary with the angle
$\vartheta_{\bm k}$ between $\bm k$ and the $x$-axis . The angle
appears as linear combination of first- and third order harmonics,
i.e. as combinations of $ \cos{\vartheta_{\bm k}}$,
$\sin{\vartheta_{\bm k}}$ and $ \cos{3 \vartheta_{\bm k}}$,
$\sin{3 \vartheta_{\bm k}}$ terms (see for
instance~\cite{Knap1996p3912,book}). The spin galvanic effect, on
the other hand, is only related to the first order harmonics in
the Fourier expansion of the non-equilibrium electron distribution
function~\cite{book}. Hence a cubic contribution leaves - for our
photocurrent measurements - the form of the Hamiltonian unchanged
(though it modifies the
spin-splitting~\cite{Jusserand95p4707,Miller03p076807,book,Lusakowski2003})
but only renormalizes the Dresselhaus constant $\beta$: The
coefficient $\beta = \gamma \langle k_z^2 \rangle$ describing
${\bm k}$-linear terms should be replaced by $\beta = \gamma
(\langle k_z^2 \rangle - k^2/4)$. Here $\gamma$ is the bulk
spin-orbit constant and $\langle k_z^2 \rangle$ is the averaged
squared wavevector in the growth direction~(see for
instance~\cite{Knap1996p3912,book}).

In conclusion, we have shown that photocurrent measurements
provide a new way to extract direct information on the different
contributions to spin orbit coupling. While we demonstrated the
potential of the method using the spin-galvanic effect in InAs
quantum wells we note that the same quantitative results were
obtained for $\alpha/\beta$  by employing the circular
photogalvanic effect~\cite{PRL01}. In contrast to the spin
galvanic effect, which requires an in-plane magnetic field to
prepare the necessary in-plane spin-orientation
{\boldmath$S_\parallel$}, the latter experiment is carried out at
zero magnetic field. The method can also be used for other
material systems like GaAs quantum wells, where, instead of the
isotropic Elliot-Yafet spin-relaxation mechanism, the anisotropic
D'yakonov-Perel mechanism dominates. In this case the anisotropy
of the spin-relaxation process~\cite{Averkiev02pR271}, which
results in an anisotropic spin distribution
{\boldmath$S_\parallel$}, must be taken into account.

Acknowledgements: We thank T. Dietl for helpful discussion. This
work is supported by the DFG, RFBR, INTAS, ``Dynasty'' Foundation
--- ICFPM, RAS and Russian Ministries of
Science and Education.

\end{document}